\documentclass[aps,prb,twocolumn,floatfix,longbibliography,showpacs,nofootinbib,10pt]{revtex4-2}

\usepackage{natbib}
\usepackage[utf8]{inputenc}
\usepackage{graphicx}
\usepackage{dcolumn}
\usepackage{bm}
\usepackage{amsfonts}
\usepackage{amsmath}
\usepackage{amssymb}
\usepackage{braket}
\usepackage{color,soul}
\usepackage{wasysym}
\usepackage{mathrsfs}
\usepackage{times}
\usepackage[dvipsnames,svgnames,table]{xcolor}
\usepackage{soul}
\usepackage{hyperref}
\usepackage{fancyhdr}
\usepackage[english]{babel}
\usepackage{longtable}
\usepackage{multirow}
\usepackage{float}
\usepackage{comment}
\usepackage{makecell}
\usepackage{enumitem}

\hypersetup{
    unicode=true,			
    pdftoolbar=true,			
    pdfmenubar=true,			
    pdffitwindow=false,			
    pdfstartview={FitH},		
    pdfauthor={FDR},			
    colorlinks=true,			
    linkcolor=NavyBlue,			
    citecolor=Maroon,			
    filecolor=NavyBlue,			
    urlcolor=NavyBlue       		
}

\newcommand{\hbm}[1]{\hat{\bm{#1}}}

\begin{document}

\preprint{APS/123-QED}

\title{
      Engineering Floquet moir\'e patterns for scalable photocurrents
      }

\author{Hernán L. Calvo}
\affiliation{Instituto de Física Enrique Gaviola (CONICET) and FaMAF, Universidad Nacional de Córdoba, Argentina}
 
\author{Luis E. F. Foa Torres}
\affiliation{Departamento de Física, Facultad de Ciencias Físicas y Matemáticas, Universidad de Chile, 8370448, Santiago, Chile}

\author{Matias Berdakin}
\email{matiasberdakin@unc.edu.ar}
\affiliation{Consejo Nacional de Investigaciones Científicas y Técnicas (CONICET), Instituto de Investigaciones en Fisicoquímica de Córdoba (INFIQC), X5000HUA, Córdoba, Argentina}
\affiliation{Universidad Nacional de Córdoba, Facultad de Ciencias Químicas, Departamento de Química Teórica y Computacional, X5000HUA, Córdoba, Argentina}
\affiliation{Universidad Nacional de Córdoba, Centro Láser de Ciencias Moleculares, X5000HUA, Córdoba, Argentina}

\begin{abstract}
While intense laser irradiation and moiré engineering have independently proven powerful for tuning material properties on demand in condensed matter physics, their combination remains unexplored. Here we exploit tilted laser illumination to create spatially modulated light-matter interactions, leading to two striking phenomena in graphene. First, using two lasers tilted along the same axis, we create a quasi-1D supercell hosting a network of Floquet topological states that generate controllable and scalable photocurrents spanning the entire irradiated region. Second, by tilting lasers along orthogonal axes, we establish a 2D polarization moiré pattern giving rise to closed orbital propagation of Floquet states, reminiscent of bulk Landau states. These features, imprinted in the bulk of the irradiated region and controlled through laser wavelength and tilt angles, establish a new way for engineering quantum states through spatially modulated light-matter coupling.

\end{abstract}

\maketitle

\section{\label{sec:intro} Introduction}

The quest to manipulate material properties on demand drives much of modern research in materials science and condensed matter physics~\cite{basov2017, noauthor2020}. Two particularly promising approaches have emerged: dynamic control through light-matter interactions~\cite{bao2022} and structural control through spatial modulation~\cite{andrei2020}. While each approach has shown remarkable success independently, the combination of these approaches, though largely unexplored, offers unprecedented control over material properties. Such combination can be achieved in two ways: by illuminating twisted structures~\cite{rodriguez2021,vogl2023}, or, as we do here, by generating space-dependent modulation through carefully tuned lasers.

Strong coupling with light has proven particularly powerful, with theoretical~\cite{lopez2008,syzranov2008,oka2009,kibis2010,calvo2011} and experimental~\cite{mahmood2016,wang2013} studies demonstrating gap opening with hybrid Floquet-Bloch states. Recent breakthroughs, including direct observation of these states in graphene~\cite{merboldt2024,choi2024,liu2024} and beyond Dirac materials~\cite{zhou2023}, and the modulation of nonlinear optical response in 2D materials~\cite{shan2021,kobayashi2023}, have validated theoretical predictions and opened new frontiers in light-based material control~\cite{bao2022}. Perhaps most striking is the ability to induce Floquet topological phases in otherwise trivial materials~\cite{oka2009,lindner2011,rudner2020}, characterized by non-trivial Chern numbers and robust edge states~\cite{piskunow2014,sentef2015,nag2021,ke2024}. The recent observation of light-induced anomalous Hall effect~\cite{mciver2020} exemplifies their far-reaching consequences, while various electron-photon interaction mechanisms enable spin-polarized photocurrents in both 3D~\cite{mciver2012} and 2D topological insulators~\cite{berdakin2021}, and control of quantum Hall states~\cite{lopez2015,huaman2021}.

\begin{figure}
\includegraphics[width=\columnwidth]{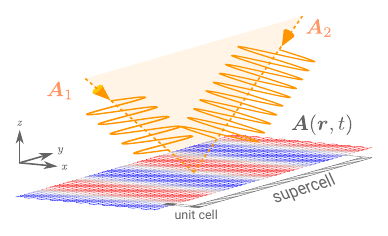}
  \caption{Irradiation scheme: Two tilted lasers create an interference pattern on a 2D material. This pattern defines a new commensurate length, outlining a supercell. The electron-photon interaction varies locally based on the position in the supercell.}
  \label{fig:axes}
\end{figure}

Here we demonstrate that combining these approaches through tilted laser illumination creates spatially modulated light-matter interactions with unprecedented control capabilities. Using graphene as a model system and experimental setups similar to those in 2D optoelectronics~\cite{mciver2020,liu2024}, we reveal two novel scenarios: 
\begin{enumerate}
\item Irradiation with two lasers tilted along the same axis produces a quasi-1D supercell, resulting in a network of Floquet topological states that span the entire irradiated region. Unlike the edge-confined states observed in previous studies~\cite{mciver2020}, these bulk topological states generate controllable and scalable photocurrents whose magnitude can be tuned by adjusting the lasers' tilt.
\item Tilting lasers along orthogonal axes generates a 2D polarization moiré pattern, creating alternating regions of laser polarization. This configuration gives rise to closed orbital propagation of Floquet states, reminiscent of bulk states in quantum Hall systems.
\end{enumerate}

These findings establish a new way of engineering quantum states through spatially modulated light-matter coupling. In the following, after introducing the theoretical framework, we build our understanding progressively: from the known case of perpendicular illumination to the emergence of bulk topological states under tilted illumination, culminating in the creation of 2D moiré patterns and their implications for controlling quantum states in materials.

\section{\label{sec:method} Semiclassical description for laser fields}

The effect of the laser on the electronic properties of the system will be described through a semiclassical approximation, as the considered average number of photons is sufficiently large. The electric field $\bm{E}$ enters as a time-periodic term in the Hamiltonian, with period $\tau=2\pi/\omega$. This will be included through the gauge $\bm{E}(\bm{r},t) = -\partial_t \bm{A}(\bm{r},t)$, where $\bm{A}$ is the vector potential. We begin analyzing the effect of tilting the laser incidence with respect to a 2D graphene layer, placed along the $xy$ plane. The situation is schematically shown in Fig.~\ref{fig:axes}, where we consider two laser beams, whose incidence direction is not perpendicular to the layer, but inclined at an angle $\pm\theta$ to the $z$ axis and along the $yz$ plane. To proceed, we first consider a single laser with perpendicular incidence along the $z$ axis, such that the vector potential reads:
\begin{equation}
\bm{A}_\perp(\bm{r},t) = \text{Re}[\bm{A}_0 e^{i(\bm{\kappa} \cdot \bm{r}-\omega t)}],
\end{equation}
where $\bm{A}_0 = A_0 (\hbm{e}_x + e^{i\phi}\hbm{e}_y)/\sqrt{2}$, with $A_0$ the laser intensity and $\phi$ the polarization, being $\phi=0$ ($\pi/2$) for linear (circular) polarization, respectively. The wave vector of the beam is $\bm{\kappa} = -\kappa \hbm{e}_z$, and its magnitude is $\kappa=2\pi/\lambda = \omega / c$, with $c$ the speed of light. The vector potential for the tilted laser is obtained from a rotation in an angle $\theta$ around the $x$ axis, yielding:
\begin{equation}
\bm{A}(\bm{r},t) = \frac{A_0}{\sqrt{2}}[\cos(\alpha-\omega t) \hbm{e}_x
+\cos(\alpha-\omega t+\phi)\!\cos\theta \hbm{e}_y],
\label{eq:Arotado1}
\end{equation}
where $\alpha = \alpha(y) = \kappa y \sin\theta$ accounts for the spatial variation along the lattice. Therefore, the introduction of a tilted incidence of the laser generates a spatial dependence of the vector potential, yielding a supercell structure that is drawn by the commensuration wavelength of the electric field (see Fig.~\ref{fig:axes}). The commensuration condition $\alpha(L) = 2\pi$ is obtained for supercell widths ($L$) that satisfy 
\begin{equation}
L=\frac{\lambda}{\sin\theta},
\label{eq:con1}
\end{equation}
providing a handle to manipulate $L$ with experimental parameters. For armchair nanoribbons (see Appendix~\ref{sec:uc}) the commensuration wavelength can thus be expressed in terms of the repetition of $N$ unit cells in the $y$ direction (see Fig.~\ref{fig:axes}) as: 
\begin{equation}
L=\sqrt{3}Na \rightarrow N = \frac{\lambda}{\sqrt{3}a\sin\theta}.
\label{eq:con2}
\end{equation}
Although Eq.~(\ref{eq:Arotado1}) is instructional to show how tilting the laser incidence induces a spatial dependence in the vector potential through $\alpha$, with a single laser this spatial dependence manifests only as a phase shift along the $y$-direction, while the polarization remains uniform throughout the supercell (see Appendix~\ref{sec:singlelaser}). In our simulations, both this phase modulation and the reduction of the $y$-component by $\cos\theta$ are fully included. However, since a single tilted laser does not create polarization interfaces within the supercell, no new topological features emerge beyond those already known from uniform illumination.

\begin{figure*}
\includegraphics[width=\textwidth]{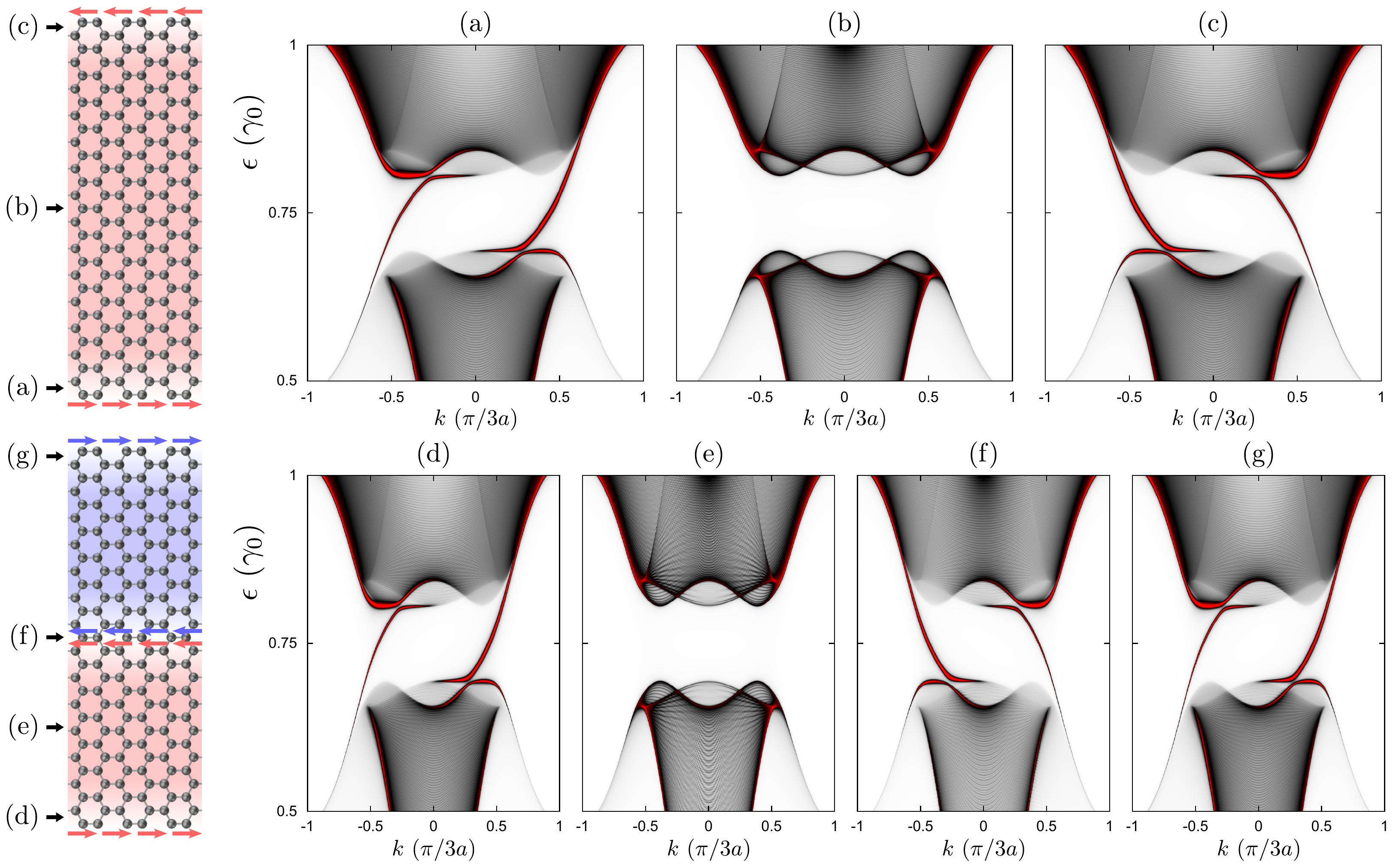}
  \caption{Perpendicular irradiation. (a-c) Graphene nanoribbon irradiated with a circularly right-handed polarized laser (red spot). The panels show the LDOS, on the logarithmic scale $\log(1+\rho)$, at the depicted positions of the ribbon (black arrows) and the colored arrows highlight the FES/FIS propagation direction. (d-g) Graphene nanoribbon irradiated with two lasers with opposite polarizations (red and blue spots for right and left-handed polarization, respectively) creating a polarization interface. We used $\hbar\omega =1.5 \, \gamma_0$ and $\zeta = 0.1$ in the two cases.}
  \label{fig:perperndicular}
\end{figure*}

As we shall see below, in order to obtain non-trivial effects from the spatial dependence of the vector potential, we extend the above considerations to the inclusion of a second laser of the same frequency and with incidence angle $-\theta$, such that the commensuration condition in Eq.~(\ref{eq:con1}) is preserved. In Appendix~\ref{sec:floquet} we derive the Floquet-Bloch Hamiltonian employed for the calculation of the time-averaged local density of states (LDOS) and transmission amplitudes.

\section{\label{sec:results} From edge to bulk states}

\subsection{\label{sec:perp_laser} Preliminaries: Edge states under perpendicular illumination}

To understand the non-perturbative effect of multiple and oblique laser irradiation on the electronic structure of graphene, we build on the outcomes of perpendicular illumination. Figure \ref{fig:perperndicular}(a-c) shows the case of an armchair ribbon irradiated with a circularly right-handed polarized laser. The time-average local densities of states (see Appendix~\ref{sec:floquet}) at the edges of the ribbon (a,c) and at the bulk of the irradiated region (b) are depicted separately. It is known that for the case of graphene, to open a gap hosting Floquet topological states requires breaking of time-reversal symmetry~\cite{oka2009}, as provided by the circularly polarized irradiation. A Chern number characterizes the bulk of Floquet topological insulators (FTI)~\cite{lindner2011}, that behave as insulators, while topologically protected states, called Floquet edge states (FES), propagate at the physical edges of the sample, where the Chern number changes. A particularity of the FTI is that they present two gaps, one of them at the Floquet zone boundary (ZB, $\epsilon=\pm \hbar\omega/2$) and the other one at the Floquet zone center (ZC, $\epsilon=0$). As in the regime of small laser intensity ($\zeta$) the ZB gap is larger than that of the ZC~\cite{calvo2011}, we will focus on the first during the rest of the work. Panels (a) and (c) show the LDOS at each edge of the sample, where it can be readily seen that two states bridging the ZB gap appear in each figure. It has been demonstrated that these states are localized at the edges and are robust to perturbations~\cite{piskunow2014}. Besides, these states are chiral as they possess opposite group velocities, and this chirality is controlled by the handedness of the laser. The existence of FES was proposed theoretically~\cite{oka2009,lindner2011,kitagawa2010}, and observed experimentally~\cite{rechtsman2013,mciver2020}.

\begin{figure*}
\includegraphics[width=\textwidth]{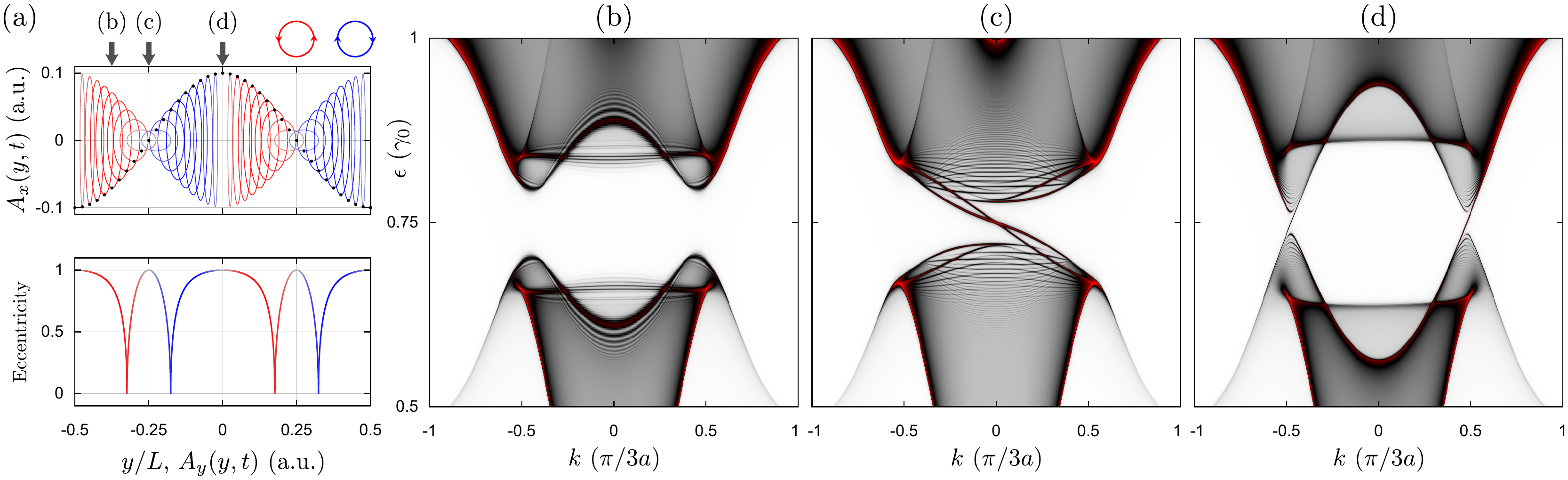}
  \caption{(a) Vector potential (top panel) and polarization eccentricity (bottom panel) profiles along the $y$ direction of the supercell for a cycle $\tau$ of the tilted irradiation described in Eq.~(\ref{eq:Atilted1}). The colors denote left (blue) and right (red) polarizations, respectively. The black dots in the top panel show the vector field values for $t=0$. (b-d) LDOS, on the logarithmic scale $\log(1+\rho)$, within the bulk of the irradiated region at positions highlighted with arrows in (a). The incidence angles are $\theta_1=-\theta_2=\pi/3$, and we considered linear polarizations, such that $\phi_1 = 0$, and $\phi_2=\pi$. In addition, we used $\zeta_1=\zeta_2=0.1$ and $\hbar\omega=1.5 \, \gamma_0$. With this set of parameters $L \approx 1400 \, l_0$, with $l_0=\sqrt{3}a$ the length of the unit cell.}
  \label{fig:tiltedLDOS}
\end{figure*}
Floquet topological states can appear in different conditions. For example, in Ref.~[\onlinecite{calvo2015}] the authors study the surface of a 3D topological insulator irradiated with two lasers with opposite handedness. They found that, as a result of the inversion of the Chern number in each irradiated region, the chirality of FES is also inverted and a pair of co-propagating states appear at the interface between the lasers. These \textit{Floquet interface states} (FIS) have also been described for semimetals under similar irradiation conditions~\cite{firoz2019,bonasera2022}. Figure~\ref{fig:perperndicular}(d-g) shows that similar results are obtained when a ribbon is irradiated with two lasers with opposite polarizations. Panels (d) and (g) show that states propagating at the physical edges of the sample have the same group velocity, unlike what can be seen in panels (a) and (c). The bulk of each irradiated region is gapped (see panel e), and at the interface of the laser spots, two pairs\footnote{Although the LDOS shows two peaks crossing the ZB gap, these correspond to valley degenerated states, as in armchair ribbons the two valleys fall in the $\Gamma$ point.} of co-propagating FIS appear, with opposite velocities to those localized at the edges. 

The above results are in good agreement with Ref.~[\onlinecite{calvo2015}] and bulk-boundary correspondence arguments,
 and establish the fundamental connection between laser polarization and the emergence of Floquet topological states. A key insight is that interfaces between regions of opposite laser polarization host co-propagating states with well-defined chirality. Building on this understanding, we now explore how tilted illumination can create controllable patterns of such interfaces throughout the bulk of the material, rather than just at the physical boundaries or at a single interface between laser spots.

\subsection{\label{sec:tilted_laser} Engineering polarization interfaces through laser interference} 

As it was mentioned in Sec.~\ref{sec:method}, tilted irradiation with a single laser can be understood as a simple modification of the field eccentricity due to its projection on the lattice. To obtain nontrivial effects, we consider the
inclusion of a second laser of the same frequency and  opposite incidence angle ($-\theta$). In addition, we will consider linear polarizations, such that $\phi = 0$ for the laser tilted in $\theta$ and $\phi=\pi$ for the second one tilted in $-\theta$, respectively. Under these conditions the resulting vector  potential reads:
\begin{equation}
\bm{A}(\bm{r},t) = \sqrt{2}A_0 \left[ g_x \cos \omega t \, \hbm{e}_x+g_y \sin\omega t \, \hbm{e}_y \right],
\label{eq:Atilted1}
\end{equation}
where $g_x = \cos\alpha$ and $g_y = \sin\alpha\cos\theta$, and we used the same intensity for the two lasers. This exhibits a periodic pattern along the $y$ direction as shown in the top panel of Fig.~\ref{fig:tiltedLDOS}(a), maintaining translational invariance in the $x$ direction. In this figure, the ellipse drawn by the resulting field in one period is shown at discrete positions of the supercell, and the color encodes the resulting polarization handedness.

It can be seen that the interference between both lasers gives rise to a space-dependent eccentricity $\varepsilon$\footnote{Considering an elliptical polarization, we define its eccentricity as $\varepsilon= (1-r_\text{min}^2/r_\text{max}^2)^{1/2}$, with $r_\text{max}$ and $r_\text{min}$ being the semi-major and semi-minor axes of the ellipse, respectively.} that results in an alternation of the polarization handedness and total intensity. For this configuration, polarization frontiers can be observed at $nL/4$ (with integer $n$). There are two important features to highlight from this result: 1) The frontiers are not equal, as can be seen in the bottom panel of Fig.~\ref{fig:tiltedLDOS}(a) where $\varepsilon$ of the resulting field is shown. The rate of change of $\varepsilon$ near the frontiers is smaller for $n=\{0,\pm 2\}$ than the the corresponding for $n=\pm 1$. Besides, these two groups of interfaces also differ in their polarization, since they are aligned along the $x$ and $y$ directions, respectively. 2) Although the commensuration is guaranteed, the resulting vector potential breaks inversion symmetry along the supercell direction. The phase and amplitude of $\bm{A}$ match at the supercell's boundaries, but an inversion of the $y$ coordinate does not preserve the handedness of the vector potential. These new flavors of the resulting field are central to the creation of a network of FIS throughout the irradiated region, which is the main outcome of this work, and would explain the appearance of zero-bias photocurrents, which we discuss in the next section.

To understand the effect of tilted irradiation on the LDOS we build on from the discussion of Fig.~\ref{fig:perperndicular} and start by showing the LDOS at discrete positions of a supercell in the bulk of the irradiated region (i.e., without physical boundaries). The selected positions are indicated with arrows in Fig.~\ref{fig:tiltedLDOS}(a). The first position (b) is chosen between two interfaces and the LDOS is expected to behave like the bulk of an irradiated graphene layer, e.g., panels (b) or (e) of Fig.~\ref{fig:perperndicular}. Accordingly, in the LDOS shown in Fig.~\ref{fig:tiltedLDOS}(b) we observe a bandgap opening at $\hbar\omega/2$ as it is expected for the bulk of a FTI. In position (c) the resulting vector potential is linearly polarized but an abrupt change in polarization can be observed nearby. A small shift on the $y$ coordinate leads to a perfectly circularly polarized field with left- or right-handed direction, akin to the irradiation condition of Fig.~\ref{fig:perperndicular}(f). Thus, this switch of the handedness implies a change in the Chern number, and in the LDOS two doubly degenerate FIS can be observed. In the position highlighted with (d) the situation is similar, so in analogy to what is observed for Fig.~\ref{fig:perperndicular}(f) we expect to obtain FIS propagating in the opposite direction than those in Fig.~\ref{fig:tiltedLDOS}(c). This is indeed what we find but with two important differences: The energy gap is smaller, and the $k$-dipersion of these FIS is narrower (note that states in Fig.~\ref{fig:tiltedLDOS}(d) do not traverse the full gap region in the $k$ direction). For details on how the local properties of the resulting vector potential in each interface modifies these features refer to Appendix~\ref{sec:softhard}.
\begin{figure*}
\includegraphics[width=0.9\textwidth]{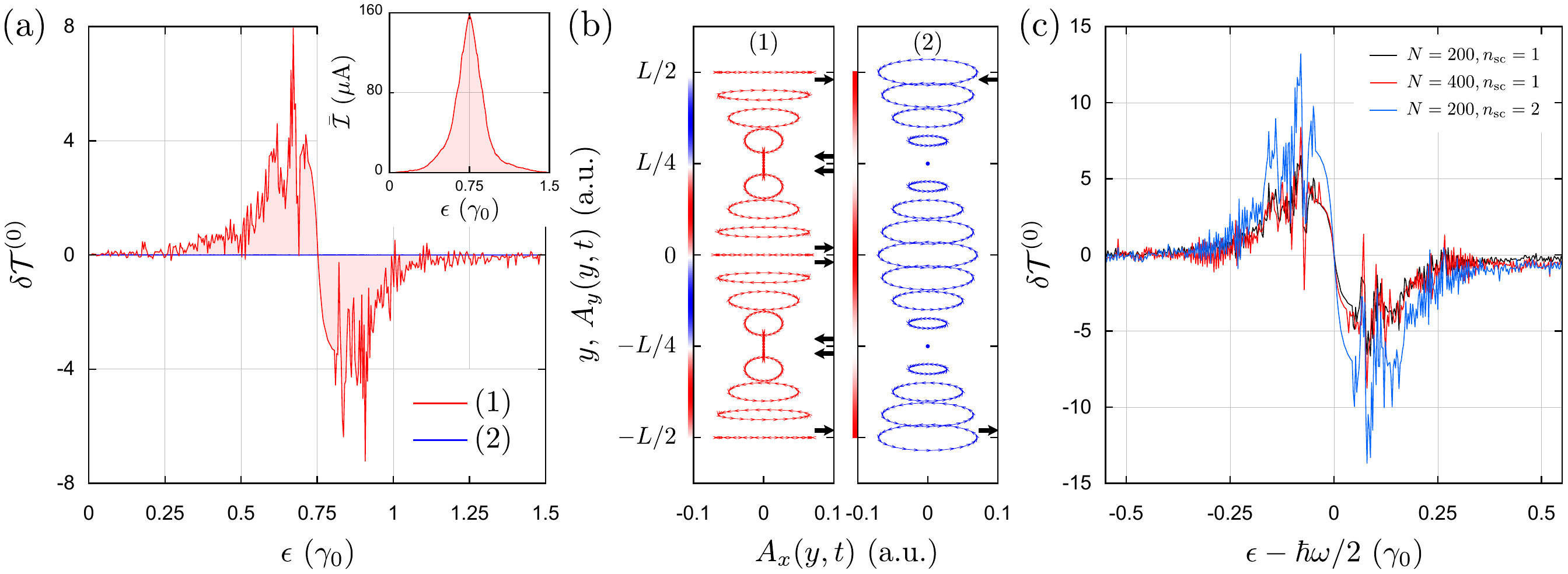}
  \caption{Laser interference and photocurrents. (a) Time-averaged difference of the transmission coefficients $\delta \mathcal{T}^{(0)}$ for the configurations shown in panel b. Inset: pumping current around the ZB. (b) Vector potential profiles for two different configurations: (1) $(\phi_1,\phi_2)=(0,\pi)$ (red), (2) $(\pi/2,\pi/2)$ (blue). In all cases we use $\theta_1 = -\theta_2 = \pi/3$. The color bars at the left of each panel denote the polarization handedness (right-handed in red and left-handed in blue), while the arrows at the right of each panel indicate edge/interface states. Due to valley degeneracy, each arrow counts two chiral states. (c) Scaling of $\delta\mathcal{T}^{(0)}$ with the number of interfaces. Here we use $W=L$ and $N=200$ (black), $W=L$ and $N=400$ (red), $W=2L$ and $N=200$ (light blue). The other parameters are: $\hbar \omega = 1.5 \, \gamma_0$ and $\zeta=0.2$.}
  \label{fig:trans}
\end{figure*}

\subsection{Scalable photocurrent generation} 

Although up to this point the discussion of the results obtained for the tilted irradiation can be built on top of what it is known for the generation of FIS in the frontier of two lasers with opposite handedness, it is worth highlighting that the overall behavior of the system with space dependent coupling  is different from what is known for perpendicular irradiation, at least in two central points:\\

\noindent {\textit{1. Topological bulk states}}.-- In most cases, conventional FES only appear at the \textit{physical} boundaries of the sample. In contrast, the Floquet topological states described in Fig.~\ref{fig:tiltedLDOS} are a bulk property that is repeated throughout the irradiated region. Note that frontiers of different polarization handedness and their corresponding topological states are inherent to the bulk of the system. For the configuration shown in Fig.~\ref{fig:tiltedLDOS}, each supercell contains four polarization interfaces. In addition, for the ZB gap, the difference between the Chern numbers at each side of the interface is $\pm 2$ per valley~\cite{calvo2015}. This topological invariant counts the difference in the number of left and right moving edge/interface states. Therefore, a supercell would contain 16 FIS crossing the ZB gap, giving rise to a manifold of topological states spanning through the whole illuminated sample. For example, the light-induced anomalous Hall effect in graphene~\cite{mciver2020} was measured using a laser wavelength of $\lambda = 6.5$ \textmu m and a laser spot of $W = 80$ \textmu m. Under these conditions, an upper limit can be achieved when $\theta = \pi/2$, such that $L = \lambda$ and the number of supercells would be $n_\text{sc} = W/L \approx 12$. Considering that each supercell contributes 16 FIS, approximately 192 FIS are obtained in the entire sample. Clearly, this number can increase if smaller wavelengths are employed. It is important to emphasize that this upper limit, achieved by minimizing $L$, corresponds to an irradiation condition that does not result in the formation of a supercell pattern. This limit is used solely to establish a feasible upper bound for the number of FIS that can fit within a laser spot under already employed experimental conditions.\\

\noindent \textit{2. Photocurrents}.-- As a smoking gun for the laser generation of a manifold of chiral states, transport signatures in solid-state devices may prove crucial. In the case of perpendicular illumination this has already been demonstrated in previous works~\cite{foatorres2014,mciver2020}. Furthermore, in the context of the present work, these types of measurements would also provide experimental evidence for the occurrence of zero-bias photocurrents.
For perpendicular irradiation, although time-reversal symmetry (TRS) can be broken using a circularly polarized laser, no photocurrents can be achieved due to the presence of inversion symmetry (IS) in clean graphene samples. As we discussed before, the resulting vector potential shown in Fig.~\ref{fig:tiltedLDOS} breaks both TRS and IS, and therefore it is expected that zero-bias photocurrents can be obtained in graphene under spatially dependent laser polarization. Although in Appendix~\ref{sec:singlelaser} we show two contributions from the vector potential to the IS breaking, here on we only refer to the one produced by the polarization pattern.

To show the emergence of a zero-bias photocurrent, we consider a typical transport setup where the interference pattern of the vector potential develops on a finite region (the sample), which is coupled to non-illuminated electronic reservoirs modeled by semi-infinite continuations of the ribbon. A non-zero photocurrent can thus be evidenced through the imbalance $\delta \mathcal{T} = \mathcal{T}_\text{RL}-\mathcal{T}_\text{LR}$ between the transmission coefficients, which can be obtained through the Floquet-Green's functions~\cite{kohler2005,foatorres2014}:
\begin{equation}
 \mathcal{T}_{\alpha \beta}^{(n)} = 4 \sum_m \text{tr} [ \bm{\Gamma}^{(m)}_\alpha \bm{G}_{\alpha\beta} \bm{\Gamma}^{(n)}_\beta \bm{G}_{\beta \alpha}^\dag],
 \label{eq:fisher-lee}
\end{equation}
where $\alpha, \beta = \{\text{L},\text{R} \}$, $\bm{G}_{\alpha\beta}$ is the retarded Floquet-Green's function, and $\bm{\Gamma}^{(n)}_\alpha$ sets the escape rate at the $n$-th replica of the $\alpha$-lead. The time-averaged photocurrent is related with the $0$-th component of the transmission coefficients, i.e., $\bar{\mathcal{I}} = 2e/h \int \delta \mathcal{T}^{(0)}(\epsilon) f(\epsilon) \text{d}\epsilon$, with $f(\epsilon)$ the Fermi-Dirac distribution function. Thus $\delta \mathcal{T}^{(0)}$ allows us to witness the appearance of a finite photocurrent. For more details on the transport simulations please refer to Appendix~\ref{sec:suptrans}. 

In Fig.~\ref{fig:trans}(a) we show the time-averaged imbalance between the transmission coefficients for the two configurations depicted in Fig.~\ref{fig:trans}(b), and for energies around the ZB. This was done for a ribbon whose width $W$ consists of a single supercell of size $L = \sqrt{3} N a$, with $N=200$ unit cells, using replicas $n=\{-1,0,1\}$.\footnote{In order to reduce the number of unit cells per supercell, we numerically release the physical relation between the laser's wavelength and frequency, taking them as two independent parameters.} 
Configuration (1) coincides with that of Eq.~(\ref{eq:Atilted1}), while in configuration (2) we use circular polarization for the two lasers, i.e., $\phi_1=\phi_2=\pi/2$. The latter preserves IS, and the polarization direction remains constant along the supercell, so the are no interfaces defined for this pattern. There are, however, two points along the supercell where the vector potential vanishes, defining a zero-gap region with dispersionless states on it, as we show in Appendix~\ref{sec:nointer}.

The simultaneous breaking of TRS and IS, present in (1), produces a non-zero $\delta \mathcal{T}^{(0)}$ which displays an antisymmetric behavior around $\hbar\omega/2$. We can see a smooth increase of $|\delta \mathcal{T}^{(0)}|$ within the smaller gap region ($|\epsilon-\hbar\omega| \lesssim 0.04 \, \gamma_0$), and it reaches maximum values for energies in between the two gaps that are characterized by the two types of interfaces. Finally, for energies above the larger gap, the transmission imbalance decays. It is interesting to note that the observed imbalance only appears around the ZB. This can be attributed to the fact that the ZC only involves elastic electron-photon transitions, i.e., the laser-induced renormalization of the $n=0$ replica is due to a photon emission process followed by a photon absorption (or vice-versa). In consequence, the transmission imbalance is not expected in this region, at least in a two-terminal setup and in the low intensity limit $\zeta \ll 1$~\cite{buttiker2006}. For configuration (2), as expected, the recovery of IS implies a negligible $\delta \mathcal{T}^{(0)}$ in the full energy range.

Remarkably, this imbalance scales with the number of interfaces. To test this, in Fig.~\ref{fig:trans}(c) (red curve) we increased the size of the supercell to $N=400$ while keeping configuration (1) for the interference pattern, thus the number of interfaces remains the same as in Fig.~\ref{fig:trans}(a). This results in a similar imbalance as the one shown in panel (a), corresponding to $N=200$ [solid black in (c)]. On the other hand, the light blue curve corresponds to the case where the sample is composed by two supercells of size $N=200$. This implies that the number of interfaces (including physical edges) ascends from 5 to 9, which coincides with the scale factor between the two curves. In general, given the number $n_\text{sc} = W/L$ of supercells that compose the sample, the transmission imbalance scales as $(4 n_\text{sc}+1)/5$. The same is expected for the photocurrent: taking into account that this reaches a value of ${\sim}160$ \textmu A per supercell [see inset in Fig.~\ref{fig:trans}(a)],  for the estimated values of $W = 80$ \textmu m and $\lambda = 6.5$ \textmu m discussed before, a photocurrent of ${\sim}1.6$ mA would be obtained around the ZB. Note that this result provides an experimental knob to optimize the photocurrents, i.e., decreasing $\lambda$ or increasing $\theta$ will reduce $L$ with a concomitant increase of $n_\text{sc}$ and therefore, the photocurrent intensity.

For more details on the emergence of zero-bias photocurrents, their link with geometrical edges, and the effect of disorder, see Appendix~\ref{sec:replicas}.

\subsection{\label{sec:2d-moire} Two-dimensional moiré engineering}

Up to this point, we have dealt with interference patterns generated by the irradiation of two lasers tilted at opposite angles on the same axis, leading to quasi-1D supercells. In this section, we show that the interaction between the space-dependent polarization and the geometry of the lattice can lead to 2D moiré-like patterns.

\begin{figure}
\includegraphics[width=0.8\columnwidth]{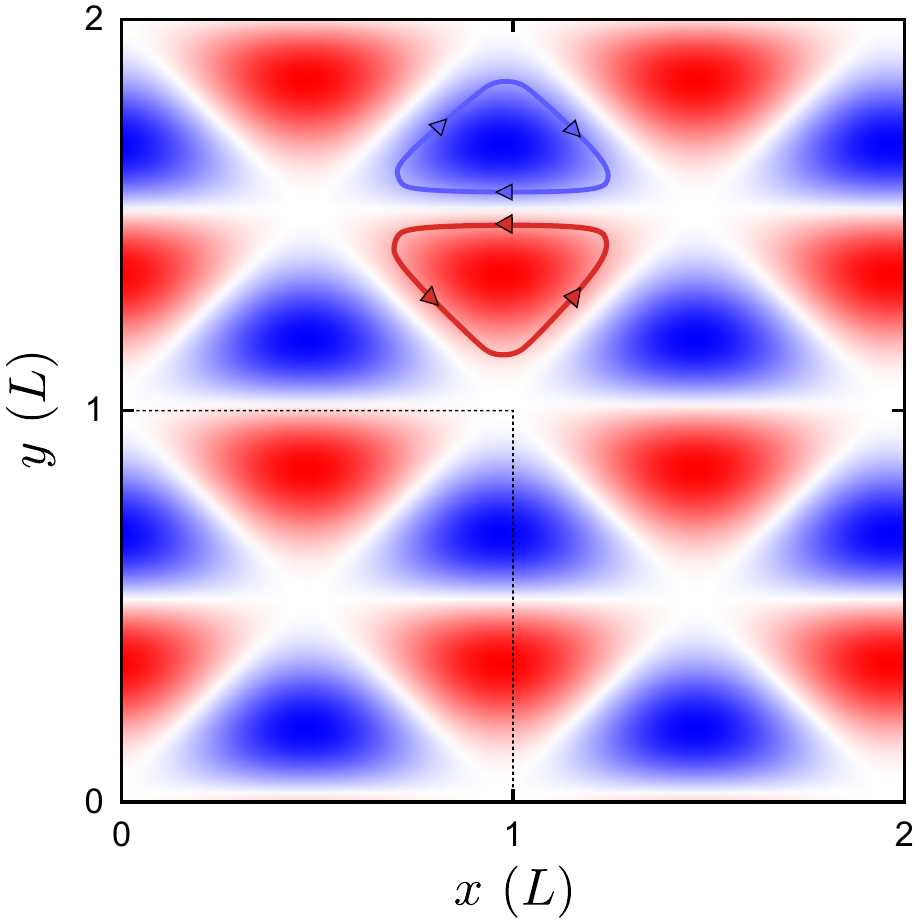}
  \caption{Polarization moiré-like patterns. Vector potential polarization along a $2\times2$ supercell. The square in dotted lines encloses a single supercell. The colors denote left (blue) and right (red) polarizations, respectively. Presumed FIS are schematically depicted with colored arrows. The irradiation parameters for both lasers tilted on the $x$ axis are $\theta_1=-\theta_2 = \pi/3$, and $(\phi_1,\phi_2) = (0,\pi)$, while for the laser tilted on the $y$ axis the parameters are $\theta_3=\pi/2$, and $\phi_3= \pi$.}
  \label{fig:2D}
\end{figure}

Equations described in Sec.~\ref{sec:tilted_laser} can be used to obtain the resulting vector potential for any tilting condition, provided the proper rotation is applied to each individual contribution.
An interesting configuration results from the addition of a third laser with a tilt on the $y$ coordinate to the irradiation condition already employed in Fig.~\ref{fig:tiltedLDOS}. In Fig.~\ref{fig:2D} we show the vector potential polarization pattern in a $2\times2$ supercell plot. It is clearly seen that a 2D moiré-like pattern is obtained under these conditions. Interestingly, each region of polarization is enclosed by the opposite polarization handedness. This picture can be read intuitively in light of what has been discussed above, supported by bulk-boundary correspondence arguments. One possibility is that the FIS circulates in closed loops in each region akin to bulk Landau states, as is schematically depicted with arrows in Fig.~\ref{fig:2D}. If so, the bulk of the irradiated region would be imprinted with dispersionless topologically protected Floquet states. However, this outcome would depend on the spacing between polarization lobes. If this is smaller than the spatial confinement of FIS this would allow them to hop to a neighboring region with the same polarization. These results also contribute to the broader context of moiré-induced topology in 2D materials~\cite{fumega2023,bennett2023,hu2023,li2024,huaman2024}.

\section{Conclusions and final remarks}

We have demonstrated that tilted laser illumination provides a powerful and exciting new approach to engineering quantum states in materials, combining the advantages of light-matter interactions with spatial modulation. Using graphene as a testbed, we showed that interference between tilted lasers creates spatially-dependent light-matter couplings with two striking manifestations: For lasers tilted along a single axis, we obtain a quasi-1D supercell hosting a network of Floquet topological states that generate scalable photocurrents spanning the entire irradiated region. By tilting lasers along orthogonal directions, we create a 2D polarization moiré pattern leading to closed orbital propagation of Floquet states, reminiscent of quantum Hall physics.

Our findings open several promising directions. The generation of controllable bulk photocurrents could enable novel optoelectronic devices, while light-induced moiré patterns offer a dynamic alternative to structural engineering of quantum states. Future work could explore multi-frequency laser configurations and the extension of this approach to other 2D materials, potentially enabling new ways to manipulate topological states on demand.

\section*{Acknowledgements}
M.B. acknowledges financial support from Consejo Nacional de Investigaciones Cient\'ifcas y T\'ecnicas (CONICET, PIBA 28720210100973CO), and Secretar\'ia de Ciencia y Tecnolog\'ia (SECyT-UNC) through Grant No.  33820230100101CB.

H.L.C. acknowledges financial support from
CONICET (PIP 2022-59241), and SECyT-UNC through Grant No. 33620230100363CB (PIDTA 2023 - Consolidar).

L.E.F.F.T. acknowledges financial support by ANID FONDECYT (Chile) through grant 1211038, The Abdus Salam International Center for Theoretical Physics and the Simons Foundation, and by the EU Horizon 2020 research and innovation program under the Marie-Sklodowska-Curie Grant Agreement No. 873028 (HYDROTRONICS Project).

\appendix
\section{\label{sec:uc} Unit Cell}
\begin{figure}[h]
\includegraphics[width=0.7\columnwidth]{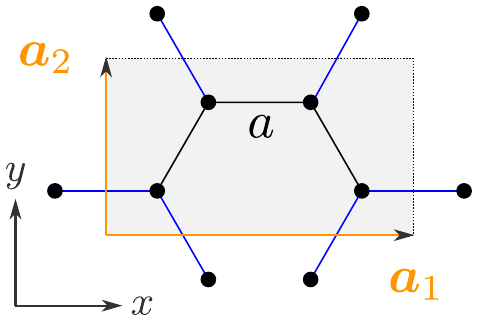}
  \caption{Unit cell as an elementary block: Schematic representation of the unit cell employed to build the irradiated region. $a$ stands for the carbon-carbon distance, $\bm{a}_1$ and $\bm{a}_2$ are the unit cell lattice vectors. Intracell hoppings are represented by black lines linking neighbor carbon atoms, while intercell hoppings are represented by blue lines.}
  \label{fig:uc}
\end{figure}

Here we present the unit cell (shaded area in Fig.~\ref{fig:uc}), composed by four carbon atoms, employed to build up the supercell of the irradiated region. The quasi-1D system is represented with an armchair nanoribbon, and the lattice vectors can be written as $\bm{a}_1=3a\hbm{e}_x$ and $\bm{a}_2=\sqrt{3}a\hbm{e}_y$, where $a = 1.42$~\r{A} is the carbon-carbon distance. Note that the final system preserves translational invariance along the $x$ direction, while it is composed by $\lambda / (\sqrt{3}a\sin\theta)$ unit cell repetitions in the $y$ direction, where $\lambda$ is the laser wavelength and $\theta$ the tilt angle.

\section{\label{sec:singlelaser} Single laser irradiation with a tilt}

\begin{figure*}
\includegraphics[width=\textwidth]{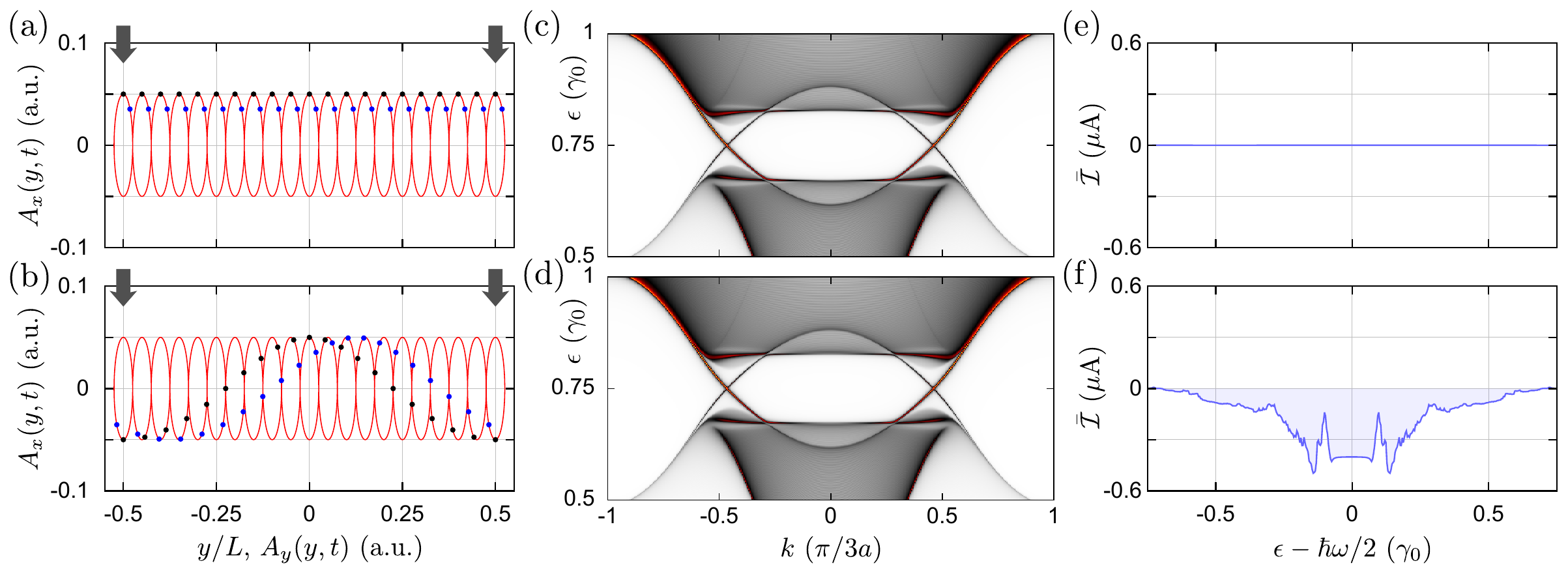}
  \caption{(a,b) Vector potential profiles along the $y$ direction traced in a cycle $\tau$ for perpendicular (a) and tilted (b) irradiation, respectively. The black and blue dots show the vector field values for $t = 0$ and $\tau/8$. (c,d) LDOS, on the logarithmic scale $\log[1+\rho(-L/2)+\rho(L/2)]$, obtained at the edges of the ribbon, highlighted with arrows in (a) and (b). 
  (e,f) Zero-bias photocurrents for configurations (a) and (b), respectively. For the configuration shown in (b), the tilt angle is $\theta=\pi/3$ and we considered circular polarization ($\phi=\pi/2$). In addition, we used $\zeta=0.2$ and $\hbar\omega=1.5 \, \gamma_0$. For configuration (a) we used the same parameters of (b), but taking $\alpha = 0$ throughout the ribbon.}
  \label{fig:singlelaser}
\end{figure*}

Here we show that, although Eq.~(\ref{eq:Arotado1}) includes a spatial dependence in the vector potential, this does not strongly 
modifies the electronic structure of the system, as compared with the case of perpendicular irradiation with an elliptically polarized laser. This is because in this case the vector potential can compactly be written as:
\begin{equation}
\bm{A}(\bm{r},t) = \frac{A_0}{\sqrt{2}} \text{Re}\left[ e^{i(\alpha-\omega t)} (\hbm{e}_x+ i \cos\theta \hbm{e}_y)\right],
\label{eq:compact}
\end{equation}
so the spatial dependence of the field only enters as a phase, with a single polarization handedness.
This is evidenced in Fig.~\ref{fig:singlelaser}, where we compare this case with the one of perpendicular irradiation with an elliptically polarized field, which would be equivalent to take Eq.~(\ref{eq:compact}) with $\alpha = 0$ along the whole supercell.

In this figure, panel (a) shows the pattern drawn by the resultant vector potential for perpendicular irradiation and elliptical polarization while panel (b) corresponds to an incidence angle of $\pi/3$ and polarization $\phi=\pi/2$. Black and blue dots in these figures denote the vector field at times $t=0$ and $\tau/8$, respectively, so the tilt of the laser manifests as the inclusion of a space-dependent phase along the $y$-direction. In the perturbative regime we consider here ($\zeta \ll 1$), this phase shift only incorporates subtle modifications of the LDOS, since we cannot distinguish them at least when evaluated at the physical edges of the sample (see panels c and d). However, the spatially dependent phase dynamically breaks the inversion symmetry of the system, since in general $\alpha-\omega t$ is not symmetric with respect to the center of the supercell at $y=0$. Interestingly, this time-dependent effect, which is not visible in the LDOS, generates an imbalance in the transmission amplitudes and, with it, a zero-bias photocurrent, as shown in panel (f). Nevertheless, regarding the discussion of photocurrents in Sec.~\ref{sec:tilted_laser} of the main text, the magnitude of this current is much smaller (and not scalable) than that obtained with two tilted lasers.

It is instructive to distinguish this single laser case with the one discussed around Eq.~(\ref{eq:Atilted1}) for two lasers, which can be written as:
\begin{equation}
\bm{A}(\bm{r},t) = \sqrt{2}A_0 \text{Re}\!\left[ e^{-i\omega t} (\cos\alpha \hbm{e}_x + i \sin\alpha \cos\theta \hbm{e}_y) \right]. 
\end{equation}
Here, we recognize a time-dependent phase ($\omega t$) times a spatially dependent vector which can invert its polarization handedness. This spatial modulation of the polarization is responsible for the interface states and scalable photocurrents discussed in the main text.

\section{\label{sec:floquet} Floquet theory}

To begin with the description of the electronic system interacting with the electric field, we consider the following nearest neighbor tight-binding Hamiltonian for a honeycomb lattice:
\begin{equation}
    \hat{H}(t) = \sum_i \epsilon_i \hat{c}_i^\dag \hat{c}_i - \sum_{\braket{i,j}} \gamma_{ij}(t) \hat{c}_i^\dag \hat{c}_j, 
\end{equation}
where $\hat{c}_i$ annihilates an electron at site $i$ of the lattice, with onsite energy $\epsilon_i$.\footnote{For graphene, we can use $\epsilon_i = 0$ for all sites.} The second sum on the right-hand side runs over nearest neighbors, connected by the hopping term $\gamma_{ij}(t)$. This term contains the time-dependent electric field, which is included through Peierls' substitution as:
\begin{equation}
    \gamma_{ij}(t) = \gamma_0 \exp{\left[i \frac{2\pi}{\Phi_0} \int_{\bm{r}_j}^{\bm{r}_i} \text{d}\bm{r}\cdot \bm{A}(\bm{r},t) \right]},
    \label{eq:hoppings}
\end{equation}
with $\Phi_0$ the magnetic flux quantum and $\gamma_0 = 2.7$ eV the bare hopping amplitude in absence of the electric field. The line integral is taken over the straight path connecting sites $j$ and $i$, with positions $\bm{r}_j$ and $\bm{r}_i$, respectively, and for the case $\lambda \gg a$ considered here we can approximate the field being constant along the bond. Through this way, it is possible to characterize the intensity of the lasers through $\zeta_\ell = 2\pi A_{0\ell}a/\sqrt{2}\Phi_0$.

The time-dependent hopping terms in Eq.~(\ref{eq:hoppings}) can be conveniently written through the Jacobi-Anger expansion as
\begin{equation}
\gamma_{ij}(t) = \gamma_0 \sum_{m,n} i^{m} \mathcal{J}_m(q_{ij}^\text{(c)}) \mathcal{J}_n(q_{ij}^\text{(s)}) e^{i(m+n)\omega t},
\label{eq:hop_t}
\end{equation}
where $\mathcal{J}_m$ is the $m$-th Bessel function of first kind and the arguments are given by:
\begin{equation}
\begin{aligned}
q_{ij}^\text{(s)} &= \sum_\ell \frac{\zeta_\ell}{a} [\Delta^x_{ij} \sin \alpha_{ij}^\ell + \Delta^y_{ij}\sin(\alpha_{ij}^\ell+\phi_\ell)\cos\theta_\ell], \notag \\
q_{ij}^\text{(c)} &= \sum_\ell \frac{\zeta_\ell}{a} [\Delta^x_{ij}\cos \alpha_{ij}^\ell + 
\Delta^y_{ij}\cos(\alpha_{ij}^\ell+\phi_\ell)\cos\theta_\ell],
\end{aligned}
\end{equation}
with $\alpha_{ij}^\ell = \kappa \bar{y}_{ij} \sin\theta_\ell$ and $\bar{y}_{ij} = (y_i+y_j)/2$, the $y$-component of the center of the bond connecting atoms $i$ and $j$. Here $\zeta_\ell = 2\pi A_{0\ell}a/\sqrt{2}\Phi_0$ sets the intensity of the lasers. Importantly, since we take nearest neighbor coupling terms, the differences $\Delta^x_{ij} = x_i-x_j$ and $\Delta^y_{ij} = y_i-y_j$ are fixed, and the $y$-dependence only enters through $\alpha$, such that the condition $L=\lambda/\sin\theta$ still holds in this case with two laser beams whose direction of propagation is defined on the same plane and with opposite angles with respect to the $z$ axis.

With the explicit time-dependence in the hopping terms of Eq.~(\ref{eq:hop_t}) we are in position to derive the Floquet Hamiltonian that allows us to move the time-periodic problem into a time-independent one. According to Floquet theorem~\cite{shirley1965,sambe1973,kohler2005}, for time-periodic Hamiltonians it is possible to find a full set of solutions to the time-dependent Schrödinger equation (TDSE). Similar to Bloch's theorem for spatially periodic systems, these can be written as $\ket{\psi(t)}=e^{-i \epsilon t/ \hbar} \ket{\varphi(t)}$, where the Floquet state $\ket{\varphi(t)}$ contains the same periodicity of the Hamiltonian, i.e., $\ket{\varphi(t+\tau)} = \ket{\varphi(t)}$, with $\tau = 2\pi/\omega$, and $\epsilon$ is a time conjugate variable known as the \textit{quasienergy}, and it is restricted to the Floquet zone $-\hbar\omega/2 \leq \epsilon \leq \hbar\omega/2$. Replacing this solution in the TDSE yields
\begin{equation}
\hat{H}_\text{F} \ket{\varphi(t)} = \epsilon \ket{\varphi(t)}, 
\end{equation}
where $\hat{H}_\text{F} = \hat{H}(t)-i\hbar\partial_t$ is the so-called Floquet Hamiltonian. Importantly, $\hat{H}_\text{F}$ can be reduced to a time-independent matrix when taking a product basis in the Floquet-Sambe space $\mathcal{F} = \mathcal{R} \otimes \mathcal{T}$. Here $\mathcal{R}$ represents the original Hilbert space and $\mathcal{T}$ is the space of time-periodic functions. The latter can be spanned by the set of orthonormal vectors $\braket{t|n} = e^{in\omega t}$, with $n$ an integer number. Working within the real space representation, a suitable basis for $\mathcal{F}$ is given by the product states $\ket{i,n} = \ket{i} \otimes \ket{n}$, accounting for the lattice site $i$ and the $n$th Fourier component, together with the inner product rule
\begin{equation}
\braket{i,n|j,m} = \int_0^\tau \frac{\text{d}t}{\tau} \braket{i|j} e^{i(m-n)\omega t} = \delta_{i,j}\delta_{n,m}.
\end{equation}
Due to its periodicity, the Floquet states $\ket{\varphi(t)}$ can be decomposed as the following Fourier series
\begin{equation}
\ket{\varphi(t)} = \sum_n e^{in\omega t} \ket{\varphi_n}
\end{equation}
and its representation in Floquet space is
\begin{equation}
\ket{\varphi} = \sum_{i,n} \varphi_{i,n} \ket{i,n},
\end{equation}
where $\varphi_{i,n}=\braket{i,n|\phi}$ is the amplitude of the Floquet state at site $i$ and Fourier harmonic $n$ (from now on, the replica $n$). Within this representation, we obtain time-independent matrix elements of the Floquet Hamiltonian, $[H_\text{F}]_{i,j}^{n,m}=\bra{i,n}\hat{H}_\text{F}\ket{j,m}$, but with an infinite number of replicas,
\begin{equation}
[H_\text{F}]_{i,j}^{n,m} = \int_0^\tau \frac{\text{d}t}{\tau} \gamma_{ij}(t) e^{i(n-m)\omega t} + n\hbar\omega \delta_{n,m} \delta_{i,j},
\end{equation}
In our case, the time-dependence comes from the hopping amplitudes given in Eq.~(\ref{eq:hop_t}), such that their Fourier transform writes:
\begin{equation}
   \gamma_{ij}^{(n)} = \gamma_0 \sum_m i^m \mathcal{J}_m(q_{ij}^\text{(c)}) \mathcal{J}_{n-m}(q_{ij}^\text{(s)}),
\label{eq:hop_n}
\end{equation}
and it represents the transition amplitude from site $j$ to site $i$ with the absorption ($n>0$) or emission ($n<0$) of $|n|$ photons.

Given the above discussion on the spatial dependence of the vector potential, we can exploit its translational invariance along the $x$ direction. In this sense, the supercell consists of a linear array of $N$ unit cells along the $y$ direction, as shown in Fig.~\ref{fig:axes}. Once this structure is established, we compute the Floquet-Bloch Hamiltonian through:
\begin{equation}
\bm{H}_k = \sum_s \bm{H}_s e^{i sk(3a)},
\label{eq:bloch}
\end{equation}
where $k$ is the electron's wave vector, restricted to the FBZ $k \in [-\pi/3a, \pi/3a)$, and the sum runs over $s = \{-1,0,1\}$. Here $\bm{H}_0$ is the Hamiltonian matrix for the supercell, and $\bm{H}_{\pm 1}$ are the coupling matrices to the neighbor supercells. These block matrices already include the photon replicas. Importantly, since we will work in the regime $\zeta \ll 1$, it is possible to safely truncate the full Floquet space by taking an adequate number of replicas, such that the observed spectrum converges.

Of particular relevance for this work is the time-averaged local density of states (LDOS), which accounts for the weight of the $k$ state at quasienergy $\epsilon$ on the site $i$ along the $y$ direction. This quantity can be obtained from:
\begin{equation}
\rho_{i,k}(\epsilon) = -\frac{1}{\pi} \lim_{\eta\rightarrow 0^+} \text{Im}\left\{[\bm{G}_k(\epsilon+i\eta)]_{i,i}^{0,0}\right\},
\label{eq:LDOS}
\end{equation}
where $\bm{G}_k$ is the Floquet-Green matrix associated with the Floquet Hamiltonian of Eq.~(\ref{eq:bloch}), i.e.,
\begin{equation}
 \bm{G}_k(\epsilon) = (\epsilon\bm{I} - \bm{H}_k)^{-1}.  
\end{equation}
Such calculation of the LDOS through the Green's function allows a fast and efficient reduction of the system's dimension (i.e., the number of unit cells), through what is known as the ``decimation'' method~\cite{calvo2013}. Following this procedure, it is possible to evaluate the LDOS in different regions of the supercell, even in a bulk situation with an infinite number of supercells on either side of the evaluation region.

\section{\label{sec:softhard} Eccentricity and polarization incidence in band gap size and FIS dispersion}

\begin{figure*}[ht]
\includegraphics[width=\textwidth]{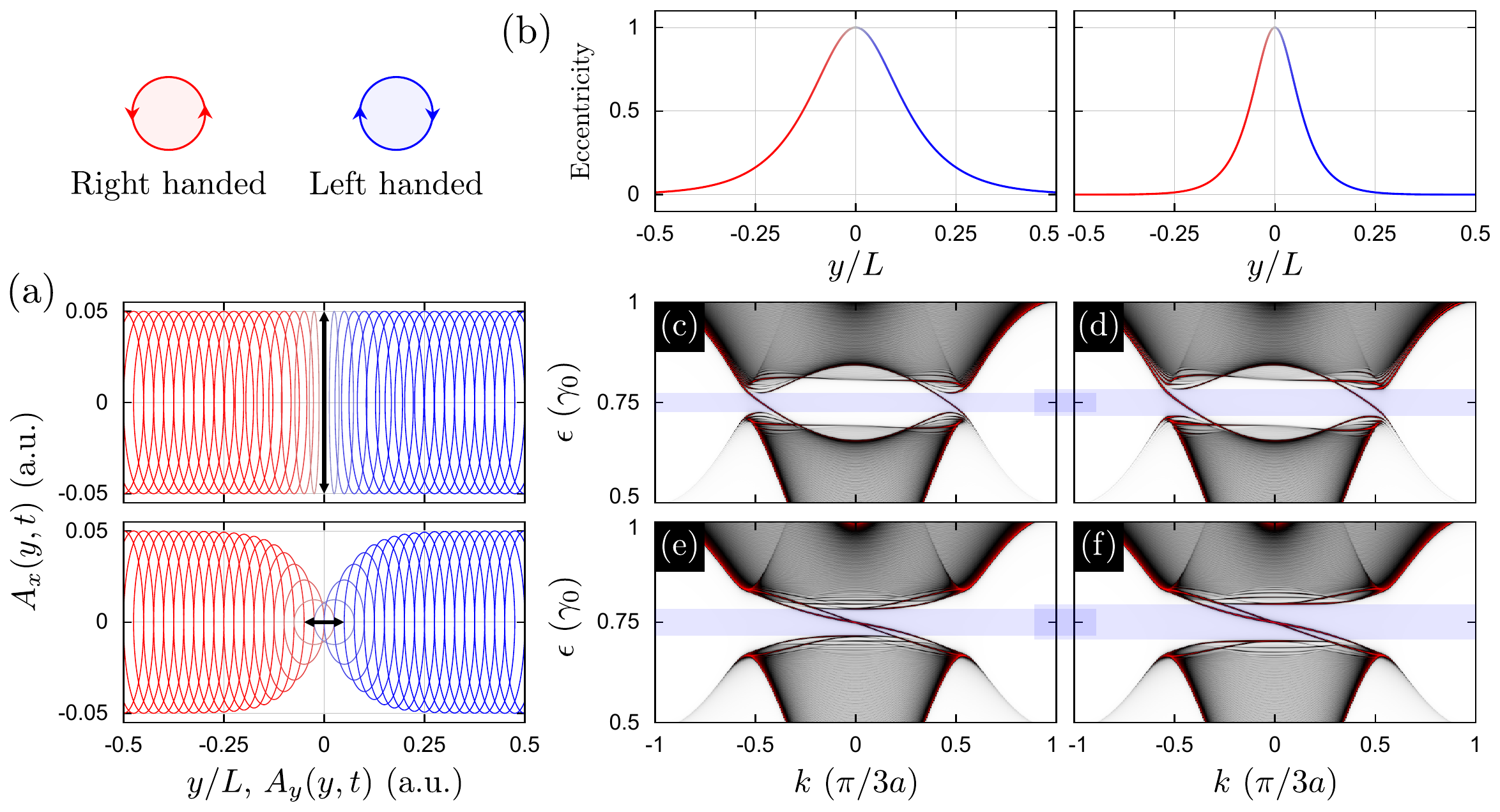}
  \caption{Local density of states, on the logarithmic scale $\log(1+\rho)$, evaluated at the interface of handedness inversion. (a) Vector potential profiles for $x$ (top) and $y$ (bottom) linear polarizations in the interface region. Red and blue lines denote right and left polarization handedness, respectively. (b) Laser eccentricity $\varepsilon$ along the supercell for $\xi = 0.1 \, L$ (left) and $\xi = 0.05 \, L$ (right), respectively. For linear polarization $\varepsilon = 1$ and for circular polarization $\varepsilon = 0$. Note that $\varepsilon$ is independent of the direction of the linear polarization at the interface. (c-f) LDOS around the ZB for the vector potential profiles in (a) and eccentricities in (b). The shaded blue areas denote the associated energy gap at the interface. The used parameters are $\hbar \omega = 1.5 \, \gamma_0$, $\zeta = 0.1$ and $L = 800 \, l_0$, with $l_0 = \sqrt{3} a$ the length of the unit cell.}
  \label{fig:softhard}
\end{figure*}

In this section, we show the effect of the local properties of the resulting vector potential at the polarization interfaces on the band gaps and the $k$-dispersion of the FIS. Here it is shown that the size of the band gap is linked to the eccentricity change rate at the interfaces, while the dispersion of the FIS inside the gap is related to the direction of polarization of the linear field. To visualize this, we consider perpendicular illumination of the sample by circularly polarized light, which is modulated in one of its components by an odd function of the $y$ coordinate, i.e.,
\begin{equation}
    \bm{A}(\bm{r},t) = A_0 \left[\cos\omega t \, \bm{e}_x + g(y)\sin\omega t  \, \bm{e}_y \right],
\end{equation}
or
\begin{equation}
    \bm{A}(\bm{r},t) = A_0 \left[ g(y)\cos\omega t \, \bm{e}_x + \sin\omega t \,  \bm{e}_y \right], 
\end{equation}
with $g(y) = - \tanh(y/\xi)$ the modulation, such that for $y = 0$ there is a change in the polarization handedness and $\bm{A}(\bm{r},t)$ is linearly polarized in either the $x$ or $y$ direction. Here $\xi$ characterizes the width of the interface region and, in turn, the rate of change of the laser eccentricity.

Figure~\ref{fig:softhard} analyzes the effect of the rate of change of the eccentricity, and the direction of polarization of the field at the interfaces, on the LDOS. Each panel (c-f) shows a LDOS that corresponds to a single condition of laser polarization at the interface [panel (a)] and a rate of change of the eccentricity [panel (b)]. For example, panels (c,d) and (e,f) show that, at a constant laser polarization at the interface [$x$ for (c,d) and $y$ for (e,f)], a higher rate of change of the eccentricity at the interface increases the band gap size. Importantly, this finds an upper bound for an ``abrupt'' interface ($\xi \rightarrow 0$), as is the case in Fig.~\ref{fig:perperndicular}(f). On the other hand, taking a given value of the eccentricity, it can be readily seen by comparing (c,e) or (d,f) that the polarization at the interface modifies the FIS dispersion. Taking into account that the interface direction coincides with the direction of the translational invariance, it can be seen that, when the linear polarization is parallel to the interface (c,d), the $k$-dispersion (or group velocity) of the FIS is larger, as the FIS only cover a reduced $k$-region of the band gap. In the second case, where the linear polarization is perpendicular to the interface (e,f), the $k$-dispersion of the FIS is smaller, since the FIS states go through the whole $k$-region spanned by the band gap. Similar results were found in Ref.~[\onlinecite{calvo2015}] for interface states at the surface of a 3D topological insulator irradiated with lasers of opposite handedness.

\section{\label{sec:suptrans} Transport simulations}

As described in the main text, the zero-bias photocurrent (also known as time-averaged pumped current) can be expressed in terms of the $0$-th component of the transmission coefficients as $\bar{\mathcal{I}} = 2e/h \int \delta \mathcal{T}^{(0)}(\epsilon) f(\epsilon) \text{d}\epsilon$, with $f(\epsilon)$ the Fermi-Dirac distribution function. For our purposes, the energy integral represents a rather elaborate task since the Floquet formalism implies an infinite number of replicas below the Fermi energy. However, a finite difference between the zeroth Fourier components of the transmission coefficients is only expected around the regions $\epsilon = 0$ and $\epsilon = \pm \hbar\omega/2$, where the $n=0$ replica mix to other replicas. The energy integral could then be taken in the vicinity of these regions, supported by the assumption that the strength of the vector potential can be taken as a small parameter, so the coupling to higher-order replicas decays sufficiently fast. Therefore, in all results shown we restrict our analysis to replicas $n=\{-1,0,1\}$, whose mixture contributes to a transmission imbalance around the ZC and ZB.

\section{\label{sec:nointer} Supercell LDOS from tilted irradiation without interfaces}
\begin{figure*}
\includegraphics[width=\textwidth]{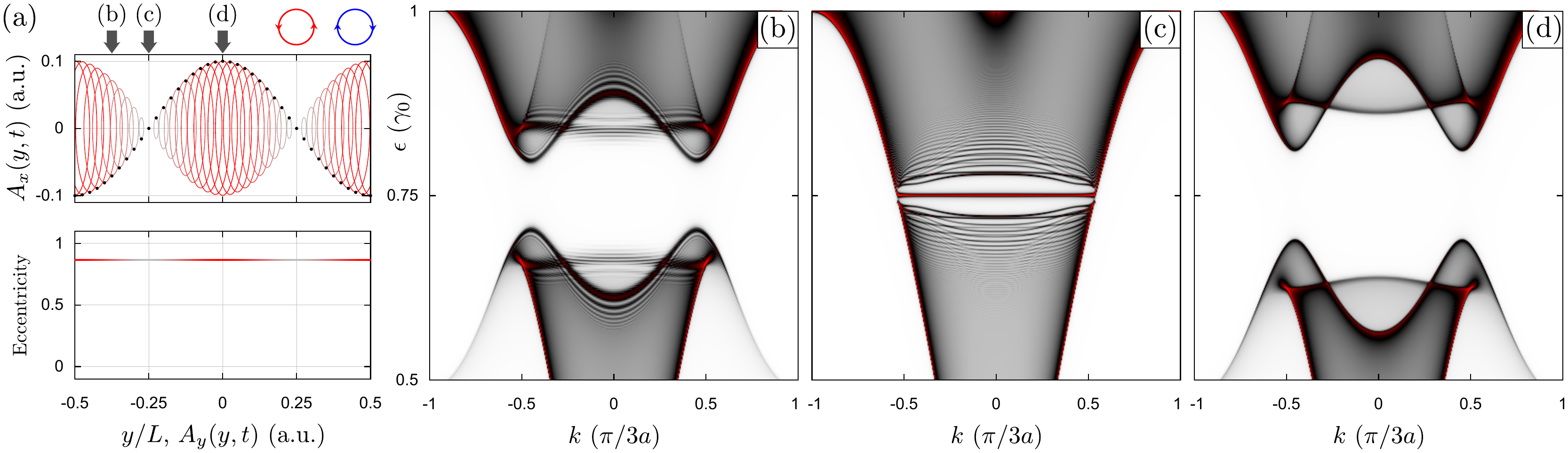}
  \caption{Vector potential (top panel) and polarization eccentricity (bottom panel) profiles along the $y$ direction of the supercell for a cycle $\tau$ of the tilted irradiation described in Eq.~(\ref{eq:conf2}). The colors denote left (blue) and right (red) polarizations, respectively. The black dots in the top panel show the vector field values for $t=0$. (b-d) LDOS obtained at discrete positions for the bulk of the irradiated region at positions highlighted with arrows in (a). Lasers parameters are : $\theta_1=-\theta_2=\pi/3$, $\phi_1 = \pi/2$, and $\phi_2=\pi/2$, $\hbar \omega = 1.5 \, \gamma_0$ and $\zeta=0.1$. }
  \label{fig:noint}
\end{figure*}

In Fig.~\ref{fig:noint} we analyze the features of the irradiation configuration (2) employed in Fig.~\ref{fig:trans}(b-2). The trajectory and the eccentricity of the resulting vector potential are depicted in panel (a), which is obtained from two circularly polarized laser beams tilted at opposite angles, yielding:
\begin{equation}
    \bm{A}(\bm{r},t) = A_0 \, g(y) \! \left( \cos \omega t \, \bm{e}_x + \sin\omega t \cos\theta \, \bm{e}_y  \right),
    \label{eq:conf2}
\end{equation}   
with $g(y) = \sqrt{2} \cos (2\pi y/L)$. As mentioned in section \ref{sec:tilted_laser}, in this configuration no polarization interfaces are produced, and the eccentricity of the vector potential remains constant along the supercell. Note that although the vector potential vanishes at $nL/4$ with $n=\pm 1$, its handedness is conserved. Therefore, there are no changes in the Chern number in these regions. The LDOS obtained at the discrete positions depicted with arrows in panel (a), which are the same as the ones employed in Fig.~\ref{fig:tiltedLDOS}(a) for configuration (1), are shown in Fig.~\ref{fig:noint}(b-d). At positions (b) and (d) the vector potential is elliptically polarized and a corresponding band gap opening is observed in their respective LDOS. In (c) the vector potential vanishes and the laser polarization is conserved. Consequently, there is no band gap opening, and a non-topologically protected flat state is observed.

\section{\label{sec:replicas} Floquet replicas, disorder and photocurrents}

In this section we discuss in more detail the connection between the occurrence of zero-bias photocurrents with spectral and transport imbalance between different Floquet replicas. In order to do so, we calculated the LDOS around the polarization interfaces of the laser pattern and weighted them according to the different Floquet replicas. A similar procedure can be done for the transmission amplitudes of Eq.~(\ref{eq:fisher-lee}).

\begin{figure*}
\includegraphics[width=\textwidth]{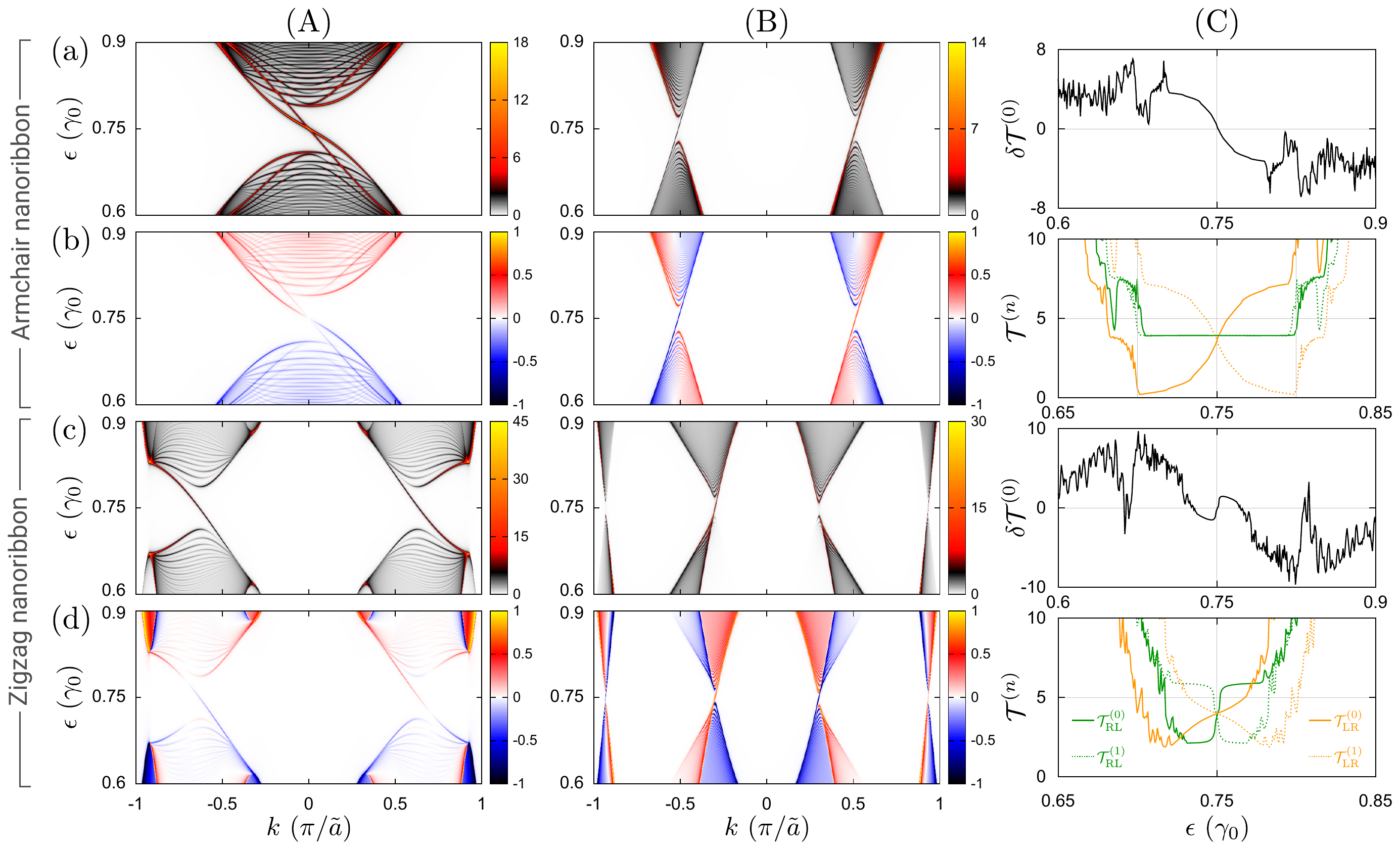}
  \caption{Replica weighted LDOS and transmission coefficients. Columns (A) and (B) show the LDOS at interfaces linearly polarized along $y$ and $x$, respectively, according to Fig.~\ref{fig:tiltedLDOS}. For these columns, rows (b) and (d) correspond to the LDOS difference of Eq.~(\ref{eq:diffLDOS}), for AGNR ($\tilde{a}=3a$) and ZGNR ($\tilde{a}=\sqrt{3}a$), respectively. Column (C) shows transmission amplitudes: (a) and (c) show the corresponding time-averaged transmission imbalance, while (b) and (d) are the transmission coefficients for the different replicas and transport direction. Green (orange) denote L$\,\to\,$R (R$\,\to\,$L) propagation, while solid and dotted lines denote the $0$-th and $1$-th replica contribution, respectively. We used $\zeta_1=\zeta_2=0.2$ and $\hbar\omega=1.5 \, \gamma_0$. For the AGNR LDOS we used $N=1400$ unit cells and for the ZGNR LDOS we used $N=800$. For the calculation of the transmission coefficients, we used a single supercell composed by $N=200$ unit cells.}
  \label{fig:replicas}
\end{figure*}

In columns (A) and (B) of Fig.~\ref{fig:replicas} we show the LDOS evaluated at the interface regions generated by the laser configuration of Eq.~(\ref{eq:Atilted1}). Column (A) corresponds to the interface with linear polarization along $y$, while column (B) shows the LDOS at the interface with linear polarization along $x$, see Fig.~\ref{fig:tiltedLDOS}. In rows (a) and (b) we use an armchair ribbon (AGNR), while in (c) and (d) we use a zigzag ribbon (ZGNR). In addition, in rows (b) and (d), columns (A) and (B), we show a normalized difference between the weighted LDOS in replicas $n=0$ and $1$, according to:
\begin{equation}
\delta_{i,k}(\epsilon) = \tanh\left( \frac{\Delta \rho_{i,k}(\epsilon)}{\xi} \right) \, ,
\label{eq:diffLDOS}
\end{equation}
with $\xi = 5$ a constant introduced to reduce contrast in the density difference, $\Delta \rho_{i,k} = \rho_{i,k}^{(0)}-\rho_{i,k}^{(1)}$, and
\begin{equation}
\rho^{(n)}_{i,k}(\epsilon) = -\frac{1}{\pi} \lim_{\eta \rightarrow 0^+} \text{Im} \{ [\bm{G}_k(\epsilon+i\eta)]^{n,n}_{i,i} \} \, ,
\end{equation}
the LDOS weighted on the $n$ replica. Finally, column (C) shows the time-averaged transmission imbalance $\delta \mathcal{T}^{(0)}$ in (a) and (c), while in (b) and (d) we decompose the transmission amplitudes with respect to its direction of propagation [$\mathcal{T}_\text{RL}$ (green) and $\mathcal{T}_\text{LR}$ (orange)] and its weight on the $n=0$ (solid) and 1 (dotted) Floquet replica.

A finite replica difference in the LDOS indicates that the associated Floquet-Bloch state $\ket{\epsilon_k}$ has different weights in the considered Floquet replicas, i.e., $\ket{\epsilon_k} = c_0 \ket{\epsilon_k,0} + c_1 \ket{\epsilon_k,1}$, with $|c_0| \neq |c_1|$. This is particularly clear for the FIS in almost all cases. Let us take for example the AGNR case. Here we can see in (A,b) that the FIS propagate from right to left and their weights change sign around $\hbar\omega/2$. For (B,b), on the other hand, the FIS propagate from left to right, and they have almost opposite weights along the whole gap region. 

An intuitive connection between these LDOS differences $\Delta \rho_{i,k}$ and the transmission amplitudes can be established by performing a $k$-integral on the Brillouin zone, such that for (A,b) this yields a finite imbalance between the replicas, while for (B,b) this integral is almost balanced. This behavior has a strong impact on the transmission amplitudes shown in panel (C,b), where a clear energy dependence appears for the R$\,\to\,$L channels. In contrast, the L$\,\to\,$R channels are almost energy independent within the gap. In addition, when summing up the two replica contributions for a given channel, e.g., $\mathcal{T}_\text{RL}^{(0)}+\mathcal{T}_\text{RL}^{(1)}$, one recovers almost quantized transmission,~\cite{kundu2013,farrell2016} whose value counts the number of Floquet interface and edge states propagating in the evaluated direction. As this value is the same for the two interfaces, the weight difference of the two $n=0$ channels yields the transmission imbalance shown in (C,a).

A similar analysis can be applied to the ZGNR, where again clear imbalance appears between the different replica contributions to both the LDOS and the transmission amplitudes. We can see in panel (C,c) that for $\epsilon \sim \hbar\omega/2$ the time-averaged transmission imbalance presents opposite signs as compared with the AGNR case of panel (C,a), meaning that such a quantity indeed depends on the edge geometry of the sample.\\

\noindent \textit{Role of disorder}.-- Another interesting point to address is the effect of disorder on the obtained photocurrents. Although a detailed calculation of the transmission amplitudes including, e.g., random vacancies through the sample is beyond the scope of this work, the presence of Floquet topological states, either located at the physical boundaries or the polarization interfaces, is robust against this type of perturbation.~\cite{piskunow2014} On the other hand, the presence of random vacancies in the ribbon naturally breaks inversion symmetry, thus contributing to the generation of zero-bias photocurrents.~\cite{foatorres2014} Therefore, we expect a competition between the above effect introduced by the laser ($\bar{\mathcal{I}}_\text{laser}$) and that given by disorder ($\bar{\mathcal{I}}_\text{dis}$). For example, in Ref.~[\onlinecite{foatorres2014}] a disorder contribution of $\bar{\mathcal{I}}_\text{dis} \sim 5$~{\textmu}A was obtained for ${\sim}0.1$\% vacancy density in illuminated graphene for a two-terminal transport setup. This means that the contribution per channel would be of $\bar{\mathcal{I}}_\text{dis}/4 \sim 1.25$~{\textmu}A. In our setup, on the other hand, the laser contribution per channel [Fig.~\ref{fig:trans}(a) inset] is $\bar{\mathcal{I}}_\text{laser}/16 \sim 10$~{\textmu}A. Adding up to this difference, while the direction of $\bar{\mathcal{I}}_\text{laser}$ is fixed through the change in the polarization handedness across the interface or physical edge, the direction of $\bar{\mathcal{I}}_\text{dis}$ may depend on the specific disorder arrangement within the interface region. In this sense, we expect no strong modifications to the photocurrents of Fig.~\ref{fig:trans} as far as the FIS are preserved.~\footnote{This implies that the localization length of the FIS should remain smaller than the separation between adjacent interfaces.} Our hypothesis rests on the fact that, by increasing the number of interfaces, the disorder contribution to the photocurrent $\bar{\mathcal{I}}_\text{dis}$ should average out since it does not have a preferential direction in each supercell, while the $\bar{\mathcal{I}}_\text{laser}$ contribution is always the same and therefore it grows with the number of interfaces. As an experimental proposal, by varying $\lambda$ or $\theta$, the number of supercells $n_\text{sc}$ could be increased, so that in a given disordered sample the expected photocurrent grows as $(4 n_\text{sc}+1)/5$, while deviations from this value due to $\bar{\mathcal{I}}_\text{dis}$ should decrease. In any case, this intriguing topic would deserve further exploration.

\bibliographystyle{apsrev4-1_title}
\bibliography{cite}

\end{document}